# Interfacial polarity control of anomalous Hall effect in $LaCoO_3$-$SrIrO_3$ heterostructures

*Yeonjae Lee, Chanyoung Shin, Youngdo Kim, Jinyoung Kim, Jaeung Lee, Miyoung Kim, Sangjae Lee*, and Changyoung Kim**

Y. Lee, Y. Kim, J. Kim, J. Lee, C. Kim
Department of Physics and Astronomy, Seoul National University, Seoul, Korea
E-mail: changyoung@snu.ac.kr

C. Shin, M. Kim
Department of Materials Science and Engineering, Seoul National University, Seoul, Korea

S. Lee
The Research Institute of Basic Science, Seoul National University, Seoul, Korea
E-mail: sangjae.lee@yale.edu

Funding: This work was supported by the National Research Foundation of Korea (NRF) grant funded by the Korean government (MSIT) (No. 2022R1A3B1077234), the Global Research Development Center (GRDC) Cooperative Hub Program through the NRF funded by the MSIT (Grant No. RS-2023-00258359), and the NRF grant funded by the MSIT (No. RS-2022-NR070586).

## Abstract text

Interfacial polarity in oxide heterostructures can drive robust charge redistribution across the interface, yet its role in tailoring topological transport phenomena like the anomalous Hall effect (AHE) remains largely unexplored. Here, we control the sign of polarity-induced charge in $SrIrO_3$-$LaCoO_3$ interfaces by independently engineering the layer stacking sequence and the atomic termination. *In-situ* angle-resolved photoemission spectroscopy directly confirms the sign reversal of the induced charge in $SrIrO_3$ through the contrasting electronic band shifts between different interfacial configurations. Magneto-transport measurements across a systematic matrix of four distinct heterostructures reveal that the AHE can be cleanly toggled on and off: it emerges exclusively under electron-doped conditions, while remaining suppressed for hole-doped $SrIrO_3$. Our results establish that this response originates from an intrinsic, Berry-curvature-driven mechanism governed by polarity-controlled band filling of $SrIrO_3$. These findings shift interfacial polarity from a passive structural consequence to an active design variable for engineering topological quantum states in oxide heterostructures.

## 1. Introduction

Oxide interfaces offer a highly versatile platform for artificially engineering physical properties, owing to the strong entanglement of charge, spin, orbital, and lattice degrees of freedom. At these heterointerfaces, several tuning parameters can be modulated independently or synergistically: charge redistribution via polar discontinuities, magnetic proximity effects, strain-induced structural distortions, and dimensional confinement. These mechanisms have enabled the realization and control of diverse emergent phenomena that are absent in bulk counterparts, including metal-insulator transitions,[1–5] the emergence of interfacial conductivity,[6–10] and topological transport.[11–15]

The 5d transition metal oxide $SrIrO_3$ (SIO) is particularly susceptible to such interfacial tuning. In SIO, strong spin-orbit coupling, Coulomb interaction, and electronic bandwidth compete on comparable energy scales, positioning its electronic and magnetic ground states near a delicate phase boundary. Consequently, even a subtle interfacial perturbation—such as proximate exchange field, charge transfer, or symmetry breaking—can drastically alter its electronic structure and magneto-transport properties. Indeed, the anomalous Hall effect (AHE) has been widely reported across various SIO-based systems, ranging from (111)-oriented systems with $SrTiO_3$ (STO),[16,17] to heterostructures with other transition-metal oxides such as $SrTaO_3$, manganites and $LaCoO_3$ (LCO).[18–27] The widespread observation of the AHE across such diverse SIO interfaces, regardless of the dimensionality or the constituent materials' magnetic character, strongly indicates that the interfacial coupling serves as a universal trigger to activate SIO's non-trivial electronic topology.

Despite these observations, a precise microscopic understanding of the emergence of AHE in SIO heterostructures remains elusive due to the complex interplay between magnetic proximity, interfacial charge transfer, and band reconstruction driven by symmetry-breaking. Notably, recent studies establish interfacial charge transfer as a dominant factor of the AHE response. Experimentally, this influence has been demonstrated through electrostatic back-gating and its thickness-dependent modulation,[21,24,26] while theoretical analyses across various 3d/5d oxide interfaces further support this picture.[28,29] While these findings establish charge transfer as a critical parameter, experimental validation through the systematic reversal of the charge transfer sign within a single material system has yet to be reported. Realizing such a sign reversal requires a strategy to deterministically invert the internal polar discontinuity at the interface, which remains a key challenge in oxide electronics.

To address this challenge, we utilize the LCO/SIO interface as an engineerable stack, exploiting LCO's dual characteristics as an electrical insulator and a polar oxide capable of driving intrinsic interfacial charge rearrangement. By precisely controlling the stacking sequence of the SIO and LCO layers and the atomic termination of the underlying STO substrate, we systematically stabilize two distinct interface configurations—$LaO/IrO_2$ and $CoO_2/SrO$—which selectively induce opposite effective charge (−q and +q, respectively) into the topologically delicate SIO layer via polar discontinuity. Magneto-transport and angle-resolved photoemission spectroscopy (ARPES) measurements across these heterostructures reveal that reversing the sign of interfacial charge transfer alone is sufficient to deterministically toggle the AHE on and off. Crucially, the AHE emerges exclusively under electron-doped conditions via an intrinsic, Berry-curvature-driven mechanism governed by polarity-controlled band filling. Ultimately, our findings demonstrate that interfacial polarity can be precisely tailored through a deliberate design of the oxide interfaces, establishing it as an active variable to engineer topological transport phenomena in oxide heterostructures.

**2. Results and Discussion**

We begin by establishing how the atomic-scale configuration governs the interfacial polarity. In $ABO_3$ perovskite, where AO and $BO_2$ layers alternate along the out-of-plane stacking direction, polarity can emerge when the individual layers are not charge-neutral. When such a polar material is interfaced with a non-polar one, effective charge can be induced at the interface to avoid polar catastrophe. Exploiting such control at the interface, reversing the stacking sequence in heterostructures composed of non-polar SIO and polar LCO changes the atomic-scale interfacial arrangement and consequently enables the reversal of the interfacial polarity. Specifically, in the LCO/SIO structure, as depicted in Figure 1a, the LaO layer with a formal charge of +1 is adjacent to SIO and is expected to induce a negative effective charge −q in SIO. In contrast, in the SIO/LCO structure shown in Figure 1b, the $CoO_2$ layer with a formal charge of −1 forms the interface and is expected to induce a positive effective charge +q in SIO. Although both structures are composed of the same oxide constituents, the polar layer in contact with SIO differs depending on the stacking sequence, leading to distinct physical properties of SIO.

To verify that the two stacking configurations indeed produce structurally different interfaces, we grew 10 unit cell (uc) LCO/10 uc SIO (LS) and 10 uc SIO/10 uc LCO (SL) heterostructures on singly $TiO_2$-terminated $SrTiO_3$ substrates and examined their atomic structures by high-resolution high-angle annular dark-field (HAADF) scanning transmission electron microscopy (STEM). As shown in Figures 1c and 1d, both LS and SL show high crystallinity and form atomically sharp interfaces. HAADF intensity profile along the stacking direction is consistent with the atomic number ($Z$)-contrast of the composing atoms, further confirming that the interface terminations follow the intended design; the LCO/SIO structure exhibits a LaO/$IrO_2$ interface (Figure 1e), whereas the SIO/LCO structure shows a SrO/$CoO_2$ interface (Figure 1f). Taken together, LCO/SIO and SIO/LCO are not merely structures with reversed stacking order, but rather a well-defined experimental platform in which the sign of the interfacial charge induced in SIO is reversed. This contrast enables us to examine how the AHE responds to the reversal of the interfacial polarity.

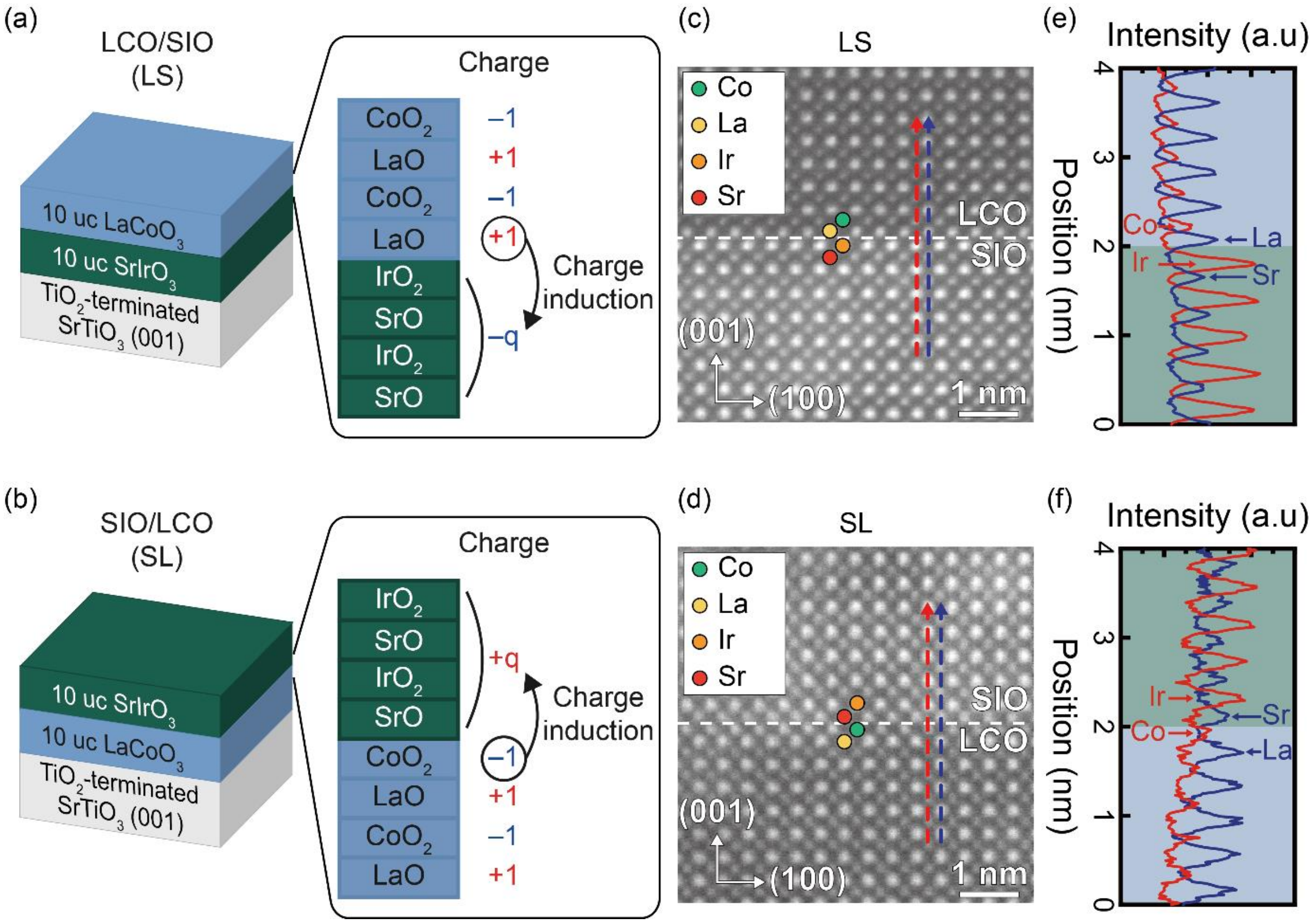

**Figure 1.** (a), (b) Schematics of the LCO/SIO (LS) and SIO/LCO (SL) heterostructures, respectively, with enlarged views of the interface region illustrating interfacial-polarity-driven charge induction in SIO. The polar LaO ($CoO_2$) layer at the LS (SL) interface induces a charge of -q (+q) to the adjacent SIO layer. (c), (d) Cross-sectional STEM images of the interfaces in the LS and SL heterostructures, respectively. (e), (f) Intensity line profiles extracted from the HAADF images along the red and blue arrows in (c) and (d), respectively. The red and blue profiles correspond to the B-site and A-site cation columns of the $ABO_3$ perovskite structure, respectively. The HAADF contrast depends on the atomic number, allowing specific A-site and B-site cations to be distinguished. The profiles confirm atomically sharp interfaces: LaO/$IrO_2$ in LS and SrO/$CoO_2$ in SL.

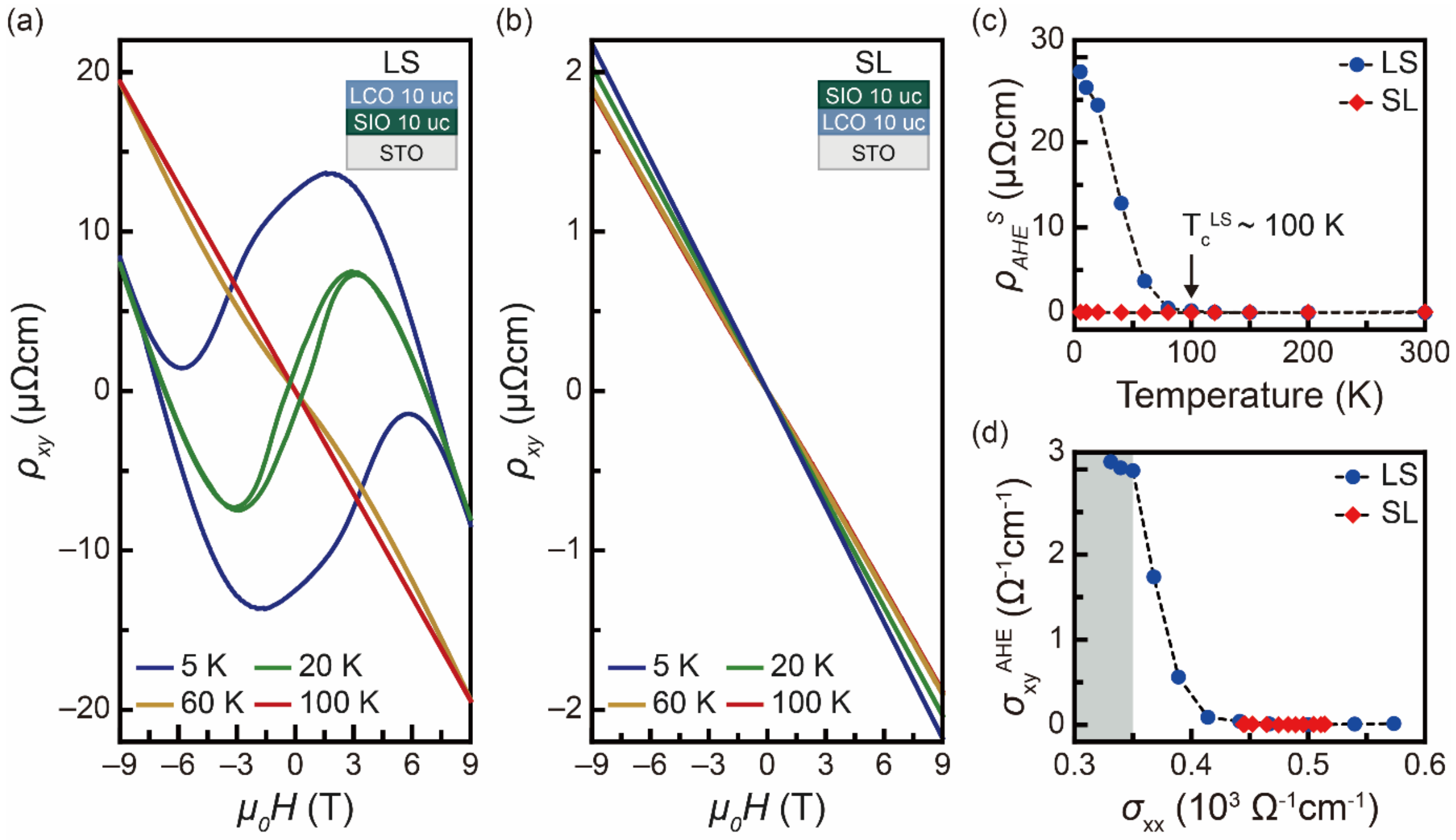

**Figure 2.** (a), (b) Hall resistivity $\rho_{xy}$ as a function of magnetic field $\mu_0$H measured at 5, 20, 60, and 100 K for the LS and SL heterostructures, respectively. The data were obtained by antisymmetrizing the raw Hall measurements. LS exhibits a clear anomalous Hall effect (AHE) with a well-defined hysteresis loop, whereas SL shows no detectable AHE. (c) Temperature dependence of the saturated anomalous Hall resistivity $\rho_{AHE}^{S}$ for LS (blue) and SL (red). LS shows a transition near $T_c^{LS}$ ~ 100 K, below which a finite $\rho_{AHE}^{S}$ develops, whereas SL shows no detectable $\rho_{AHE}^{S}$ over the entire measured temperature range. (d) Anomalous Hall conductivity $\sigma_{xy}^{AHE}$ as a function of longitudinal conductivity $\sigma_{xx}$ for both heterostructures. The shaded region highlights the plateau in LS, where $\sigma_{xy}^{AHE}$ is nearly independent of $\sigma_{xx}$, indicating the intrinsic nature of AHE in LS. SL shows no detectable AHE signal.

To investigate whether this opposite interfacial polarity affect the magneto-transport response of SIO, we performed transport measurements on LS and SL heterostructures. The magnetic-field-dependent Hall resistivity ($\rho_{xy}$) of LS exhibits a clear AHE signal with well-defined hysteresis loops, as presented in Figure 2a. As the temperature increases, the coercive field systematically decreases due to thermal activation. In contrast, SL shows no detectable AHE across the entire measured temperature range, as shown in Figure 2b. This distinction is further corroborated by the temperature-dependent saturated anomalous Hall resistivity, $\rho_{AHE}^{S}$, which reveals the onset of AHE in LS below T ~ 100 K (Figure 2c). These collective findings demonstrate that the AHE is exclusively observed in the electron-doped SIO, whereas it remains completely suppressed in the hole-doped case.

To gain further insight into the origin of the AHE, we plot the anomalous Hall conductivity, $\sigma_{xy}^{AHE}$, as a function of the longitudinal conductivity, $\sigma_{xx}$ (Figure 2d). While no measurable

AHE signal is observed for SL, LS shows a clear plateau region where $\sigma_{xy}^{AHE}$ becomes nearly independent of $\sigma_{xx}$, for $\sigma_{xx} < 350\ \Omega^{-1}cm^{-1}$ (corresponding to temperatures below ~ 20 K, highlighted in gray). This behavior strongly suggests an intrinsic origin of the AHE in LS: the emergence of AHE is fundamentally driven by the momentum-space Berry curvature associated with the topology of the electronic band structure, rather than extrinsic mechanisms like skew scattering or side-jump.[30]

Having established the significance of the interfacial atomic arrangement for the emergence of AHE, we next employ an additional tuning knob: the atomic termination of the substrate. By switching from $TiO_2$-terminated to SrO-terminated STO, we intentionally invert the interface configurations realized in the LCO/SIO and SIO/LCO heterostructures. As illustrated in Figures 3a and 3b, this termination control is expected to reverse the sign of the polarity-driven charge induced in SIO.

While $TiO_2$-terminated STO can be routinely prepared by standard substrate treatment, obtaining an SrO-terminated STO surface is not straightforward. To overcome this difficulty, we used the fact that $SrRuO_3$ (SRO) grown by PLD has SrO termination rather than $RuO_2$ termination.[31–33] By depositing SRO on STO substrate and then growing 10 unit cells of STO, we created an SrO-terminated surface for subsequent film growth (see Figure S1 and corresponding section). This template allows us to independently control the STO termination and examine its effect on the polarity-driven charge induction in SIO.

To directly verify that the sign of the induced charge in SIO changes with the interface configurations, we perform ARPES measurements to probe the electronic states of SIO/LCO heterostructures. The induced charge due to polarity is expected to reside mainly at the interface. Therefore, the SIO thickness was set to 3 uc to maintain strong sensitivity to this interfacial charge modulation. We compare three samples with 3 uc SIO as the topmost layer: S3L10, a heterostructure with 10 uc LCO beneath SIO corresponding to the SIO/LCO configuration; S3, a bare SIO film on STO serving as a reference; and R-S3L10, a heterostructure with 10 uc LCO/ 4 uc SRO/ 10 uc STO corresponding to the R-SIO/LCO configuration, where R- denotes the inclusion of the SRO layer to achieve SrO-termination. The three different films should each correspond to hole-doped, pristine, and electron-doped SIO layers, respectively.

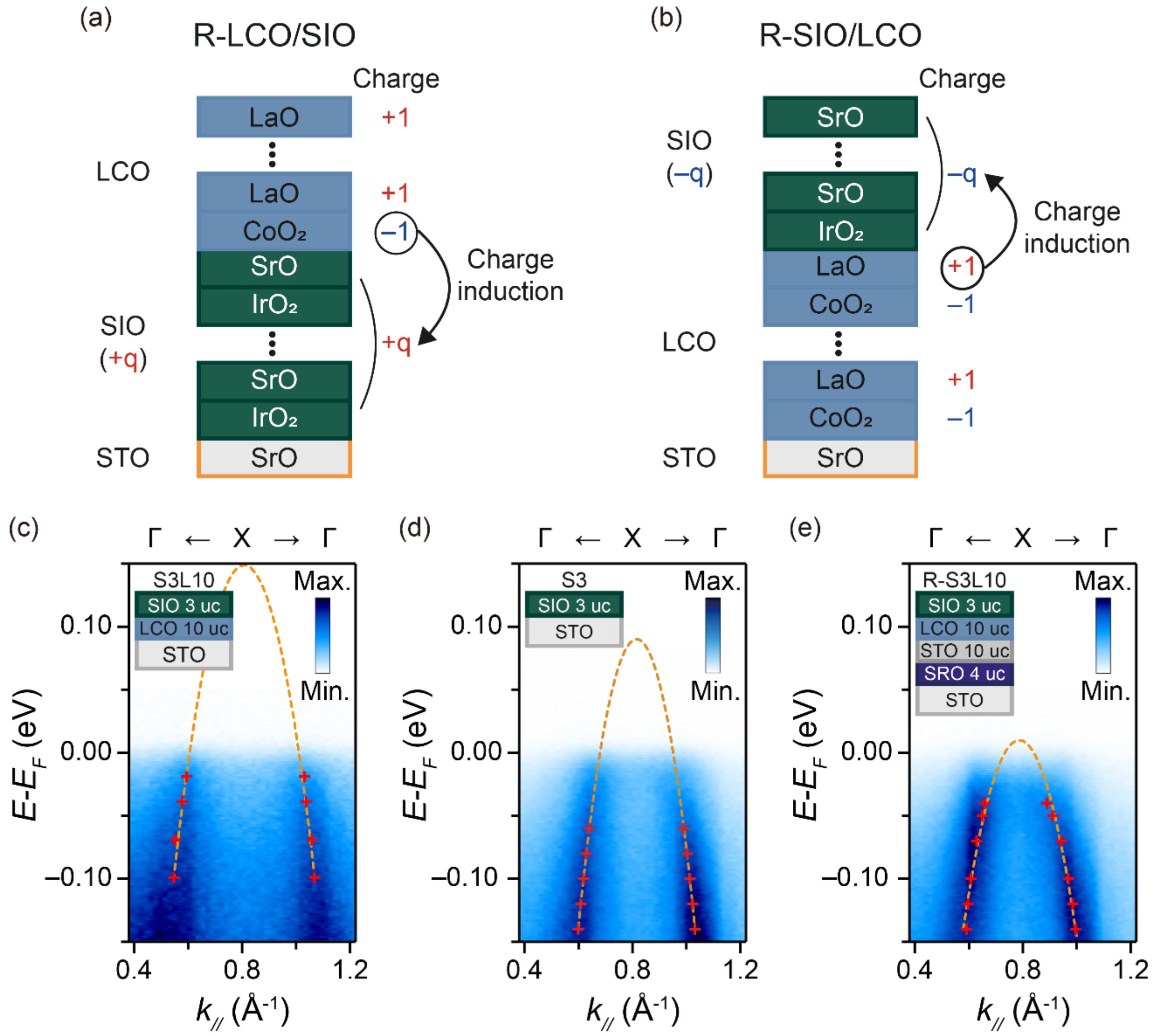

**Figure 3**. (a), (b) Schematics of interfacial-polarity-driven charge induction in SIO for the LCO/SIO (R-LCO/SIO) and SIO/LCO (R-SIO/LCO) heterostructures grown on SrO-terminated STO, respectively. Compared with the LS and SL heterostructures grown on $TiO_2$-terminated STO, the sign of the polarity-induced charge in SIO is reversed. (c)-(e) Angle-resolved photoemission spectroscopy images of S3L10, S3, and R-S3L10, respectively, measured near the X point at 5.5 K. The opposite band shifts of S3L10 and R-S3L10 relative to S3 confirm the reversal of the induced charge in SIO by the interface configuration.

Pristine SIO thin films exhibit a characteristic hole band centered at the X-point, derived from the $J_{eff}$ = 1/2 state.[34] The energetics (i.e. the valence band maximum) of this hole-band is highly sensitive to the electronic valence of Ir.[35] Tracking the evolution of the electronic structure across these samples will enable a direct evaluation of the interfacial charge transfer across different interface configurations. As shown in Figures. 3c-e, compared to the pristine SIO, the hole-band in S3L10/R-S3L10 shift towards relatively lower/higher binding energy. To quantify the exact charge transfer, we extracted $k_F$ from parabolic fits to the near-$E_F$ dispersion and estimated the carrier density change per Ir relative to pristine 3 uc SIO. The resulting charge modulation is +0.12 e/Ir for S3L10 and −0.08 e/Ir for R-S3L10. This provides direct experimental evidence that the induced charge in SIO is reversed for different interface configurations—LaO/$IrO_2$ or SrO/$CoO_2$—and further demonstrate that an engineered SrO-terminated STO template offers an independent experimental handle for tuning the interfacial polarity.

To use this additional control parameter in transport measurements, the SRO layer thickness should be kept thin so that its metallic conductivity would not contribute to the transport response of SIO. Therefore, we used a 1 uc SRO layer, whose transport contribution is negligible (see Figure S2 and corresponding section),[36] to fabricate R-LS and R-SL with interface configurations opposite to those of LS and SL (Figures. 4a,d). HAADF-STEM images and the corresponding intensity line profiles in Figures. 4b,e confirm atomically sharp interfaces, namely $CoO_2$/SrO in R-LS and $IrO_2$/LaO in R-SL, consistent with the intended interface inversion.

The Hall resistivity measurements shown in Figure 4c,f provide strong support for the central hypothesis of this work. Unlike in the original series without the SRO termination-changing layer, where AHE was observed only in LS configuration, a clear hysteretic AHE is observed here exclusively in R-SL, while R-LS shows no AHE. Thus, the stacking configuration in which AHE emerges is completely reversed when the substrate termination is changed. This result demonstrates that the emergence of AHE is not simply a consequence of placing SIO in proximity to ferromagnetic LCO, but is instead governed by the interfacial charge transfer established at the junction. As shown in Figure 4g, the $\sigma_{xy}^{AHE}$ versus $\sigma_{xx}$ plot for R-SL exhibits a clear plateau in the low-temperature regime, perfectly replicating the intrinsic scaling behavior observed in the LS heterostructure (Figure 2d).
Overall, the intrinsic AHE response emerges only in the electron-doped SIO (*i.e.* in LS and R-SL heterostructures). This is in line with theoretical predictions that SIO layers in LCO/SIO heterostructures host double Weyl points and associated band crossings that generate strong Berry curvature close to the Fermi level.[28] Although a direct quantitative comparison is not warranted since the calculation, ARPES, and transport measurements were performed on structures with different geometries and SIO thicknesses, our ARPES results nevertheless demonstrate polarity-dependent band shifts of several tens of meV. The associated shift in the relative position of the topological features with respect to the Fermi level in electron-doped SIO is responsible for the emergent AHE response.

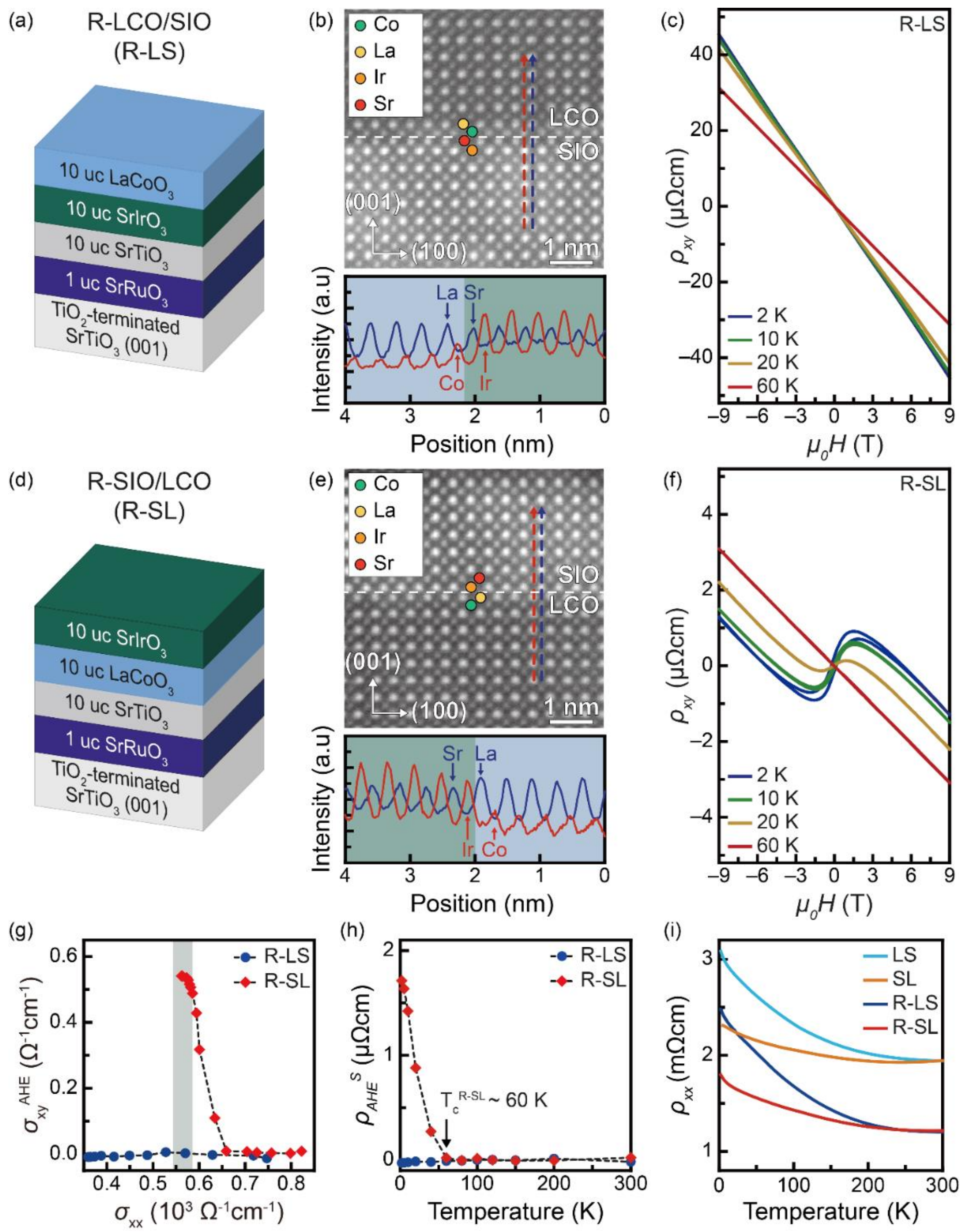

**Figure 4.** (a), (d) Schematics of the R-LCO/SIO (R-LS) and R-SIO/LCO (R-SL) heterostructures, respectively. (b), (e) Cross-sectional HR-STEM images and corresponding intensity line profiles of the LCO/SIO interface in R-LS and the SIO/LCO interface in R-SL, respectively. The profiles confirm atomically sharp interfaces, with $CoO_2$/SrO in R-LS and $IrO_2$/LaO in R-SL. (c), (f) Hall resistivity $\rho_{xy}$ as a function of magnetic field $\mu_0H$ measured at 2, 10, 20, 60 K for R-LS and R-SL, respectively. R-SL exhibits a clear AHE with a well-defined hysteresis loop, whereas R-LS shows no detectable AHE, which is the exact opposite of the trend observed in the LS and SL. (g) Anomalous Hall conductivity $\sigma_{xy}^{AHE}$ as a function of longitudinal conductivity $\sigma_{xx}$ for both heterostructures. The shaded region highlights the plateau of $\sigma_{xy}^{AHE}$ in R-SL, where $\sigma_{xy}^{AHE}$ is nearly independent of $\sigma_{xx}$, indicating the intrinsic nature of AHE in R-SL. R-LS shows no detectable AHE over the entire measured range. (h) Temperature dependence of the saturated anomalous Hall resistivity $\rho_{AHE}^{S}$ for R-LS (blue) and R-SL (red). R-SL shows a transition near $T_c^{R\text{-}SL}$ ~ 60 K, below which a finite $\rho_{AHE}^{S}$ develops, whereas R-LS shows no detectable $\rho_{AHE}^{S}$ over the entire temperature range. (i) Temperature dependence of the longitudinal resistivity $\rho_{xx}$ for LS, R-LS, SL, and R-SL, measured from 2 to 300 K.

Although R-SL and LS satisfy the same interfacial polarity condition for the emergence of AHE, the two heterostructures are not entirely equivalent in their transport characteristics. R-SL exhibits a lower transition temperature ($T_c$ ~ 60 K; Figure 4h) compared to that of LS ($T_c$ ~ 100 K). Furthermore, the longitudinal resistivity $\rho_{xx}$ reveals systematic trend across all four samples: LS-type configurations consistently show higher $\rho_{xx}$ than SL-type ones (*i.e.* LS > SL; R-LS > R-SL), while R-type heterostructures display a lower absolute $\rho_{xx}$ than their original counterparts (Figure 4i). Complementary evidence comes from the ordinary Hall effect (OHE), extracted from the linear high-field background of $\rho_{xy}$: OHE magnitude is comparable within the same stacking families (LS vs. R-LS; SL vs. R-SL), but differ by approximately one order of magnitude between LS-type and SL-type configurations. These quantitative differences collectly suggest that, beyond the sign of polarity-induced charge, the overall stacking geometry also modulates the local interfacial electrostatic environment – contributing to variations in Tc, $\rho_{xx}$, and carrier density – though the precise microscopic origin of these differences remains to be clarified. Taken together, our results establish interfacial polarity as the primary control parameter for the on/off switching of the AHE, while demonstrating that finer detailes of the transport response remain sensitive to the broader electrostatic context set by the heterostructure geometry.

**3. Summary and Outlook**

In this work, we systematically demonstrated that interfacial polarity is the key control parameter governing the emergence of the AHE in SIO-LCO heterostructures. By systematically inverting the sign of the polarity-induced charge at the SIO interface across four heterostructures—LS, SL, R-LS, and R-SL—through independent control of stacking order and substrate termination, we showed that AHE is switched on exclusively under electron-doped conditions and remains fully suppressed under hole-doping. This establishes that the mere proximity of ferromagnetic LCO to SIO is not sufficient to activate the AHE; rather it is the direction of interfacial charge transfer that determines rather that Berry-curvature contribution at the Fermi surface is enhanced. The quantitative differences in the transport properties between LS and R-SL further suggest that, while the sign of polarity-induced charge sets the primary AHE on/off condition, the broader stacking geometry may additionally modulate the local exchange environment, potentially through changes in the Co charge state and interfacial magnetic configuration.

More broadly, the ability to deterministically switch emergent magneto-transport behavior by tuning two atomic-scale growth parameters – stacking order and interface termination – establishes interfacial polarity as an active design variable rather than a passive structural consequence. This design paradigm offers a versatile platform for exploring and manipulating emergent topological phases across a wide range of strongly correlated oxide heterostructures.

## 4. Experimental Section/Methods

*Sample Growth*: SIO, LCO, SRO, and STO films were deposited on STO (001) substrates by pulsed laser deposition using a KrF excimer laser ($\lambda$ = 248 nm). Prior to deposition, the STO substrates were etched in a BOE 6:1 solution and deionized water for 30 s and 2 min, respectively, followed by annealing at 1100 °C for 1 h in air to obtain a singly $TiO_2$-terminated surface. SIO and LCO were grown at 600 °C and 100 mTorr with a laser repetition rate of 1 Hz, using laser fluences of 1.1 and 1.14 $Jcm^{-2}$, respectively. SRO was grown at 670 °C and 100 mTorr with a laser fluence of 1.9 $Jcm^{-2}$ and a repetition rate of 2 Hz, whereas STO was grown at 600 °C and 10 mTorr with a laser fluence of 1.0 $Jcm^{-2}$ and a repetition rate of 1 Hz. Samples were then cooled to room temperature at 50 °C $min^{-1}$ in 100 mTorr $O_2$. During the growth, reflection high-energy electron diffraction was used for real-time monitoring of layer-by-layer growth and thickness control.

*Transmission Electron Microscopy:* Cross-sectional scanning transmission electron microscopy (STEM) specimens were prepared using a Thermo Fisher Scientific Helios G4 focused ion beam (FIB) system. Prior to FIB milling, a protective Pt layer was deposited to minimize surface damage. High-angle annular dark-field STEM (HAADF-STEM) images were acquired using a Thermo Fisher Scientific Themis Z aberration-corrected microscope operated at 200 kV with a semi-convergence angle of 18.5 mrad.

*Electrical Transport Measurement:* Electrical transport measurements were performed using a Quantum Design Physical Property Measurement System (PPMS-9T). Prior to measurement, 60 nm thick Au electrodes were deposited in a Hall bar geometry through a metal mask using a Korea Vacuum Tech KVET-IM2002L electron-beam evaporator, and electrical contacts were formed with Al wires using a West Bond 7476D-79 wedge bonder. An out-of-plane magnetic field was applied during the measurements. Longitudinal resistance was measured from 2 to 300 K with a temperature step of 0.5 K, and Hall measurements were conducted at 2, 5, 10, 20, 40, 60, 80, 100, 120, 150, 200, and 300 K.

The anomalous Hall resistivity, $\rho_{xy}^{AHE}$, was obtained by subtracting the linear ordinary Hall component, $\rho_{xy}^{OHE}$, from $\rho_{xy}$, and $\rho_{AHE}^{S}$ was defined as the nearly saturated value at high magnetic fields. The conductivities used in the scaling analysis were calculated as $\sigma_{xy}^{AHE} = \rho_{xy}^{AHE}/[(\rho_{xy}^{AHE})^2 + (\rho_{xx}^{0})^2]$ and $\sigma_{xx} = \rho_{xx}^{0}/[(\rho_{xy}^{AHE})^2 + (\rho_{xx}^{0})^2]$, where $\rho_{xx}^{0}$ is the longitudinal resistivity at zero magnetic field.

*Angle-resolved Photoemission Spectroscopy (ARPES):* All ARPES measurements were performed in a home-built system equipped with a Scienta DA30 analyzer and a Fermi instrument He-Iα discharge lamp($h\nu$ = 21.2 eV) at a base temperature of 5.5 K. The base pressure of the ARPES chamber was $3.0\times10^{-11}$ Torr. For the *in-situ* ARPES measurements, the PLD-grown samples were transferred to a preparation chamber with a base pressure of approximately $4.0\times10^{-11}$ Torr and post-annealed at 600 °C under an oxygen pressure of $10^{-8}$ Torr for 10 min.

**Acknowledgements**

This work was supported by the National Research Foundation of Korea (NRF) grant funded by the Korean government (MSIT) (No. 2022R1A3B1077234), the Global Research Development Center (GRDC) Cooperative Hub Program through the NRF funded by the MSIT (Grant No. RS-2023- 00258359), and the NRF grant funded by the MSIT (No. RS-2022-NR070586). This work was also supported by the Institute of Applied Physics, Seoul National University. Cs-corrected STEM works were supported by the Research Institute of Advanced Materials (RIAM) in Seoul National University.

**Data Availability Statement**

The data that support the findings of this study are available from the corresponding author upon reasonable request.